\documentclass[12pt,preprint]{aastex}

\begin{document}

\slugcomment{Astrophysical Journal}

\shortauthors{Thompson, Smith, \& Hester}
\shorttitle{Star Formation in M16}

\title{Embedded Star Formation in the Eagle Nebula}

\author{Rodger I. Thompson}
\affil{Steward Observatory, University of Arizona,
    Tucson, AZ 85721}

\author{Bradford A. Smith}
\affil{University of Hawaii, Institute for Astronomy, Honolulu, HI, 96822}

\author{J. Jeff Hester}
\affil{Department of Physics and Astronomy, Arizona State 
University, Tempe, AZ 85287}

\begin{abstract}

M16=NGC 6611, the Eagle Nebula, is a well studied region of star
formation and the source of a widely recognized Hubble Space Telescope
(HST) image. High spatial resolution infrared observations with the
Near Infrared Camera and Multi-Object Spectrometer (NICMOS) on HST
reveal the detailed morphology of two embedded star formation regions
that are heavily obscured at optical wavelengths.  It is striking that
only limited portions of the visually obscured areas are opaque at 2.2
microns.  Although the optical images imply substantial columns of
material, the infrared images show only isolated clumps of dense gas
and dust. Rather than being an active factory of star production, only
a few regions are capable of sustaining current star formation. Most of
the volume in the columns may be molecular gas and dust, protected by
capstones of dense dust.

Two active regions of star formation are located at the tips of the
optical northern and central large ``elephant trunk'' features shown in
the WFPC2 images.  They are embedded in two capstones of infrared
opaque material that contains and trails behind the sources. Although the
presence of these sources was evident in previous observations at the
same and longer wavelengths, the NICMOS images provide a high
resolution picture of their morphology. Two bright stars appear at the
tip of the southern column and may be the result of recent star
formation at the top of that column. These observations suggest that
the epoch of star formation in M16 may be near its endpoint.
\end{abstract}


\section{Introduction}

Dramatic optical pictures of M16 taken with WFPC2 on HST \citep{hes96}
show three large columns or elephant trunks of molecular gas and dust
that are shaped by the stellar wind and radiation pressure from high
luminosity young stars a few arc minutes away. The columns are
designated I, II, and III going north to south.  The morphology of the
columns clearly indicates that the initial high mass star formation in
a molecular cloud region plays a large role in subsequent star
formation by altering the density structure of the gas and dust in the
surrounding region.  Large areas of low density are cleared of material
while regions of higher density form elongated structures extending
radially away from the cluster of newly formed stars.  The elongated
trunk structures are repeated on a smaller scale with ``fingers'' and
``eggs'' of material throughout the larger structure.

It has generally been assumed that new star formation is occurring
inside the trunk and egg structures and that gravitational attraction
of the protostellar condensations is retaining the material against the
action of the radiation pressure and stellar wind from the nearby young
stars.  Earlier infrared observations by \citet{hil93} indeed revealed
significant previous star formation that was not easily visible by
optical means, although they limited their quantitative analysis mainly
to objects with detectable visible flux. They determined that there
are stars up to 80 M$_{\sun}$ in the visible OB association and that
there are intermediate mass stars still above the main sequence,
assuming their distance of 2 kpc which we will adopt in our analysis.
The embedded sources discussed here are clearly visible in their image,
but were not remarked upon individually as they were not detected in
the optical image.  They also do not appear in their table of embedded
infrared sources.  The non-stellar morphology of the sources is not
clearly apparent in their published image, although there is nebulosity
near the source locations.  There is also evidence for both sources
(which we label M16ES-1 and M16ES-2 for M16 Embedded Source 1 and 2) in
the WFPC 2 F547M images.  M16ES-2,  visible as a diffuse reflection
nebulosity, was first described by \citet{hes97}.

Although at lower resolution and much brighter limiting magnitude,
M16ES-1 and 2 are clearly visible in the near infrared image of M16 by
\citet{mcc97}, however, they are not commented on in the text.  In this
black and white rendering of a portion of a much larger 3 color (J,H,K)
image the transparency of large sections of the columns is clearly
apparent.  \citet{mcc97} comments that at the limiting magnitude of
the image only 10\% of the eggs labeled by \citet{hes96} appear to
contain forming stars or chance coincidences with background sources.

Work done on the optical emission line images of the region
\citep{hes96} demonstrated the influence of photoevaporation on the
morphology of the gas and dust.  In the absence of any other effects,
photoevaporation will eventually destroy the region leaving only those
stars that managed to form before the destruction.  The observations
discussed in this paper are part of a program to study the detailed
morphology of the embedded star formation in the trunks and to study
the effects of embedded star formation on the subsequent evolution of
the region.  A detailed analysis of the sources in the region is in
preparation \citep{smth02,hes02}.

\section{Observations}

The observations utilized the NICMOS camera 3 during the HST NICMOS
camera 3 campaign of January 1998.  During the campaign the HST
secondary mirror was adjusted slightly to achieve the best camera 3
focus.  Observations were taken in columns I and II at four different
positions to cover areas also observed with WFPC2.  A fifth position in
column III was also observed  later with the higher resolution camera
2.  The camera 3 images are in the F110W, F160W, and F222M wide and
medium band filters, while the camera 2 images are in the F110W, F160W
and F205W filters.  All observations are four point dithered with a
total of 768 seconds of integration in the F222M  and F205W filters and
704 seconds in the F110W and F160W filters.  The pixel size of camera 3
is 0.2 arc seconds which gives a 51 by 51 arc second field of view for
each of the 4 camera 3 observations.  Camera 2 has a pixel size of
0.075 arc seconds for a 15 by 15 arc second field of view.

In addition to the continuum images, the position at the head of column
I was also imaged in molecular hydrogen and atomic hydrogen.  The line
and continuum narrow band filters, F187N and F190N, isolate the atomic
hydrogen Paschen $\alpha$ line while the F212N and F215N filters
isolate the H$_2$ S(1) line.  Again four dither positions were utilized
with 96 and 160 second integrations for each position in Paschen
$\alpha$ and H$_2$ respectively.

\section{Data Reduction}

The NICMOS images were reduced utilizing IDL based procedures developed
for image production in the Hubble Deep Field \citep{thm99}.  The
individual dither position images for each filter were then aligned and
the median image calculated via an additional IDL software interactive
procedure called IDP3 \citep{sto99}.  The associated continuum images
were subtracted from the line filter images to produce the final Paschen~$\alpha$ and H$_2$ emission line images.

\section{Infrared Morphology}

Figure~\ref{fig-ir} is a color image of the infrared observations with
the RGB colors corresponding to the F222M, F160W and F110W (2.2, 1.6,
and 1.1 \micron) filters respectively. This image shows that, rather
than being dense columns of gas and dust, there are only a few regions
of the elephant trunks that contain high concentrations of dust. The
heads of columns I, II and III contain dense dust capstones, however,
most of column I has a low dust column density.  The limited spatial
coverage precludes a similar statement  from our data for column II but
there is an obvious thinning of dust away from the column head.
Inspection of the \citet{mcc97} image confirms that this thinning
continues until another capstone dust region occurs to the southeast.
Below the dust region at the top of the column I there is a large
region of low dust density and then several areas of higher dust
density to the southeast. Each of these have similar geometries which
point back toward the OB cluster. The high density dust regions in both
columns have sub-structures that appear to point back to the embedded
sources at the top of the columns.  This suggests that, although the
overall geometry of the columns is controlled by the radiation and
stellar wind from the O star complex, the geometry of the smaller dense
regions may be affected by local sources.  Section~\ref{sec-loc}
discusses this possibility.

Although there is a very high density of stellar sources, not evident
in the optical images,  the darkness of the high density regions
indicates that the majority of the stellar sources lie behind the
columns.  Since the position of M16 is very close to the direction of
the galactic center, many of the fainter sources may be stars in the
galactic plane, not associated with the M16 complex.  An extensive
foreground screen of dust, not directly associated with the columns,
obscures the stars from view at optical wavelengths.

Fig.~\ref{fig-c2} shows the camera 2, higher resolution, image of the
tip of column III.  The tip of column III is concave and 2 bright stars
reside in that region.  These stars are too bright to be distant
background stars but the density of stars of similar brightness in the
images indicates that is a non-negligible chance that their location in
the concave region of the tip could be just chance coincidence.  The
possibility that the two stars recently formed in the tip of the column
is discussed in Section~\ref{sec-c3}.  Following the lead of our
nomenclature for the embedded sources we label the two stars M16S-1 and
M16S-2.

The location of high density dust regions at the leading faces of the
columns and the morphology of those regions is supportive of the idea
that shocks, driven in advance of the ionization front are important in
compressing the columns.  The location of the two embedded sources at
the very leading edge of the columns further supports this concept.
The sparsity of high density dust regions also supports the suggestion
\citep{hes96} that photoevaporation is disrupting as well as triggering
active star formation.

The lack of dense dust in the columns below the two embedded sources
and their associated dust structures is consistent with the radio,
sub-mm, and infrared maps presented by \citet{wh99} and the near
infrared images of \citet{mcc97}.  These maps and images show strong
emission at the locations of the embedded sources but very little
emission along the columns until another area of strong emission to the
southeast, beyond the extent of our images.  CO maps by \citet{wh99}
indicate that there is molecular gas along the whole length of the
columns but that its density is far less than at the location of the
sources. The combination of the near infrared images and the longer
wavelength observations show that only limited areas of the columns are
capable of sustaining current star formation, which explains the
observation of \citet{pil98} that the star formation rate in the
columns, based on the mid-infrared flux, appears to be very low.  The
eastward color gradient in the background from dark to red in
fig.~\ref{fig-ir} is due to thermal emission in the F222M band
increasing from the relatively dust free region above the columns to
the dustier area of the columns.

Fig.~\ref{fig-3colors} shows the individual infrared continuum
images along with the well known optical image. Comparison of the
infrared images with the optical indicates that most of the dust
structures observed in the optical are identifiable in the infrared
image. In particular each ``finger'' structure observed in the optical
appears to have a capstone of dense dust. The blue nebulosity near the
heads of the columns and along the edges of the dust regions in
fig.~\ref{fig-ir}  is probably emission from hydrogen Paschen~$\beta$
and possibly He I 10830.  Both of these lines are in the passband of
the F110W filter.  Two clearly non stellar amorphous sources (M16ES-1
and M16ES-2) lie at the tops of columns I and II and appear to be
emerging from the high dust density regions.  The positions of these
sources, based on the World Coordinate System (WCS) of the images are
given in Table~\ref{tab-dat}. We have not performed astrometry on the
stars in the images, so these positions are only as accurate as the
guide star positions. A typical uncertainty on guide star positions is
$\pm 1\arcsec$.

It is not clear whether the sources are clusters of stars or reflection
nebulosities created by single sources.  The lack of Paschen~$\alpha$
emission at the position of M16ES-1 indicates that it must either be a
cluster of relatively low mass stars or an object (or objects) that has
not yet reached the main sequence (Sec.~\ref{sec-lines}).  We will
assume that both sources are star formation sources, based solely on
their association with the high density dust regions.  Sections
\ref{sec-pht} and \ref{sec-msfr} discuss these sources more
thoroughly.

In addition to the two amorphous embedded sources near the tops of the
columns, at least three other point-like sources are visible at the tips
of fingers and at the top of a nearby dense dust region.  One
finger source is at the tip of a very thin finger, M16-E42
\citep{hes96}, in the southwest corner of the lower image of column I.
It is also visible in the optical image of the same area in
Fig.~\ref{fig-3colors}.  The thin finger is easily visible in the
optical and F110W images but is difficult to detect in the longer
wavelength images.  Another finger source is just to the north of the
first finger source, is at the top of the adjacent broader finger of
dust (M16-E39). This point-like source is visible in all of the
infrared images but not in the optical.  Although unlikely, we can not
rule out that it is a chance superposition of a background star given
the high density of stars in the region.  Finally there is a very
bright point source on the shoulder of a dust region just northeast of
the the very bright star in the WFPC image that falls just off the
NICMOS images.  This source is at the location of a photoionized
surface (M16-E33).  All of these sources will be discussed more
thoroughly in a future publication \citep{hes02}.

\subsection{Extinction and Density in the Columns} \label{ssec-ex}

It is useful to get an estimate of the extinction and density of the
material in column I.  Sophisticated treatments of extinction utilizing
J, H, and K near infrared colors such as NICE \citep{lad99} and NICER
\citep{lom01} have been developed. Unfortunately our images have
significant emission in two of our bands, Pa $\beta$ and He 10830 in
the F110W band and thermal emission in the F222M band which complicates
these analyses.  Instead we use a simpler technique that compares the
average magnitude of the stars in the F160W image both in and out of
the columns. We will consider two regions in column I, a region in part
of the dark capstone of column I and a mid-column region where the
column appears quite transparent.  Two reference regions above the
columns are chosen to be clear of the dust associated with the
columns.  The position of all of the regions are shown in
Fig.~\ref{fig-3colors}.  We utilized the DAOPHOT photometric analysis
program \citep{stet01} to measure the magnitudes of the stars in the
image down to a Vega magnitude of 19.  For the two regions a and b
above the columns the average magnitudes are 17.26 and 17.13 for 173
and 102 stars respectively.  In the intra-column region c the average
magnitude was 17.58 for 91 stars.  The average uncertainty in magnitude
for the regions is 0.12 magnitudes.  From this we conclude that the
intra column region has an extinction of $0.4 \pm 0.2$ magnitudes at 1.6
\micron.

If the dust in column I follows the same extinction law as the dust
near the star VI Cyg No. 12 the ratio of extinction at 1.6 $\micron$ to
V is 0.175 \citep{rie85} giving a visual extinction of 2.3 $\pm$ 1.2
magnitudes.  The empirical relation between X-ray absorption and
optical extinction \citep{sew99} gives $N_H/A_V = 1.9 \times 10^{21}$
atoms cm$^{-2}$ mag$^{-1}$.  This leads to a column density of hydrogen
in the region of (4.4 $\pm$ 2.2) $\times 10^{21}$ hydrogen atoms
cm$^{-2}$.  If we assume that the columns depth is roughly equal to its
width we get a depth of about $8 \times 10^{14}$ cm and a hydrogen
density of (5.5 $\pm$ 2.8) $\times 10^6$ H atoms cm$^{-3}$.  If the
hydrogen in primarily molecular the number density of molecules will be
1/2 this value.  This density is the same as for a high density
molecular cloud but the extinction is low because of the small column
depth.

In the dark region d of column I we do not detect any stars.  Aperture
photometry of individual stars indicates that  stars at a F160W
magnitude of 23 are readily detectable.  Even if we limit the
background stars in the dark area to the average magnitude of 17.2, the
lower limit on the extinction at 1.6 $\micron$ is 6 magnitudes.  The
extinction is most likely more than this value as there are probably
stars several magnitudes brighter than 17.2 behind the area.  Using the
same arguments as above, the visual extinction in the dark area is at
least 34 magnitudes and the hydrogen column density at least $6.5
\times 10^{22}$ H atoms cm$^{-2}$.  This is of course a lower limit and
the actual extinction and density are most likely many times the
numbers used here.

\section{Photometry and Luminosity of the Sources} \label{sec-pht}

Precision photometry of the embedded sources is difficult due to the
extended morphology and underlying emission.  We chose a circular
aperture of 2 arc seconds radius for our photometry.  Although there is
emission outside of this radius it appears that most of the flux is
inside the area. We estimate the underlying emission by averaging over
a circular annulus of inner and outer radii of 3 and 6 arc seconds
centered on the embedded source.  Both of these procedures will tend to
underestimate the flux if the source flux extends beyond the 2 arc second
radius or if source flux also lies in the annulus used for background
flux calculation.  The results of this simple photometry are given in
Table~\ref{tab-dat}.  Note that although the F160W and F222M filters
correspond closely to the usual H and K bands, the F110W filter is much
bluer than the usual J band.

We can use the mid and far infrared observations compiled by
\citet{wh99} to estimate the luminosities of the embedded sources.
Both have strong radio, SCUBA, MSX, and ISOCAM sources centered on
their locations.  To estimate the luminosities, we take the long
wavelength fluxes in Fig. 3 of \citet{wh99} and integrate under the
curves to get the long wavelength luminosity.  To estimate the shorter
wavelength luminosity we use our measured photometry and the mid-IR
fluxes from the MXS data base maintained at IPAC (M16ES-1 only).  We
then extend the mid-IR flux to 200 microns by drawing a straight line
in the log wavelength versus log flux plot to intersect the long
wavelength plot at 200 microns and again integrate under the curves.
This probably underestimates the flux between 14 $\micron$ and the
far-IR giving a lower limit on the luminosity.  The luminosities
determined in this manner are 200 (+50, -25) and 20 (+10,-5) L$_{\sun}$
for M16ES-1 and 2 respectively.  Each of these luminosities could be
for a single object or for a cluster of objects if the nonstellar
images are due to a group of forming stars as opposed to a reflection
geometry.  Unfortunately the ISOCAM fluxes of \citet{pil98} do not list
the integrated fluxes at the sources.  They only list the fluxes of
point-like objects away from the positions of the embedded sources. The
ISOCAM images from that work, presented by \citet{wh99}, clearly show
strong extended emission at the locations of both embedded sources.

\section{Morphology of the Embedded Star Formation Region} \label{sec-msfr}

M16ES-1 and M16ES-2 appear non-stellar in all of the infrared images.
M16ES-1, however, may have a strong point-like component.  The F110W
image in Fig.~\ref{fig-3colors} clearly shows a dust bar that runs across
the location of M16ES-1.  The main component of M16ES-1 in the F222M
image is at the western edge of the bar.  The peak of emission is
almost point-like with an extension to the northwest.  This hints that
the extended nature of the emission may be primarily due to reflection
off nearby dust structures. In all filters there is nebulosity to
the north of M16ES-1 which is most likely a combination of reflected
light from M16ES-1 and the emission from gas ionized by the O stars.
All three of the NICMOS continuum filters contain emission lines in
their bandpasses.  M16ES-2 is less extended with nebulosity extending
toward the east.  The intensity of the nebulosity relative to the
source increases with decreasing wavelength.  \citet{hes97} suggests
that M16ES-2 is currently emerging from its surrounding dust cloud via
photoevaporation.  At a low stretch the image of M16ES-2 appears to
be double with surrounding nebulosity.  Polarimetry with a revived
NICMOS will be able to separate a point source or sources from the
reflected components.

\section{Emission Line Morphology} \label{sec-lines}

Fig.~\ref{fig-lines} shows the emission in hydrogen
Paschen~$\alpha$, hydrogen  H$\alpha$, the H$_2$ S(1) line at
2.12 \micron, and the 2.2 $\micron$ continuum at the position of
M16ES-1.  The other locations do not have line images.  These figures
show the locations of the ionized and molecular gas as well as the area
of high dust concentration.  The H$_2$ image shows the extent of the
molecular gas which conforms well with the dust regions shown by the
2.2 $\micron$ continuum image.  The exception is the area just above
the embedded source where the structure seen in the H$_2$ image is
absent.  This is probably due to hydrogen Brackett $\gamma$ emission,
which falls in the F222M filter band, hiding the structure.  The strong
atomic hydrogen emission above the embedded source (in the direction of
the O star cluster) is due to a local HII region ionized by the
radiation from the O star cluster.  The ionized gas is the result of
photo-evaporation of gas from the head of the column as described in
\citet{hes96}, although it is possible that M16ES-1 may also have a role
in moving gas into this ionized region.  Both the H$\alpha$ and the
Paschen $\alpha$ images show emission where the ionizing radiation from
the cluster strikes the surface of the dust and molecular cloud. The
mechanisms for this surface emission are detailed by \citet{hes96}.

The absence of Paschen $\alpha$ emission at the location of M16ES-1
indicates that it is not a source of ionizing radiation. This
observation rules out M16ES-1 as a single ZAMS star. At a luminosity
of $200 L_{\sun}$ it would be a B6-B7 star on the main sequence
producing an HII region with a $\log{N_{e}^2V} = 53.9$ \citep{th84}.
This would produce detectable flux in Paschen~$\alpha$.  Its
absence indicates that either the source is made up of multiple, lower
luminosity objects or that it is an object or objects that has not yet
reached the ZAMS.  We cannot comment similarly on M16ES-2 as there is
no Paschen~$\alpha$ image for that source.

Above the embedded source the morphology of the H$_2$ emission is quite
different from the ionized hydrogen emission.  The lack of molecular
material in this region confirms that most of the gas in the area has
been ionized by the O star cluster stars.  Below the embedded source the
H$_2$ emission follows the same structure as the atomic hydrogen
emission, defining the surface of the dust and molecular cloud.  This
suggests that the H$_2$ emission may be due to photo-excitation
\citep{bl87} rather than by shock excitation.  Unfortunately, with only
an image in one line, we can not discriminate between the two emission
processes, however, near infrared Fabry-Perot images of the heads of
the columns \citep{all99} indicate that the majority of the H$_2$
emission is from fluorescent photo-excitation.

\section{Local Source Geometry} \label{sec-loc}

The infrared images of the dust clouds associated with M16ES-1 and 2
show dust tendrils that point directly toward the embedded
sources.  This strongly suggests that, although the general structure of
the M16 columns is determined by the O star cluster, local dust and
molecular structures are being influenced by the embedded sources.  The
concentration of the dense dust capstones into a cone of approximately
60 degrees in width on the side away from the O star cluster attests to
the strong influence of the cluster radiation on the dust structure. To
assess the roles of each, we can make some elementary estimations based
on the relative luminosities and distances of the O star cluster and
the embedded sources.  Taking the cluster luminosity as $2 \times 10^6
L_{\sun}$ from the O star luminosity and the M16ES-1 luminosity as
$200 L_{\sun}$, along with a distance between the cluster and the source
of 2 parsecs, we find that at distances less than 0.02 parsecs the
radiation pressure from the embedded source exceeds that from the
cluster.  Here we have assumed that the absorption efficiency of the
gas and dust is independent of the spectrum of the radiating source.
At the distance of M16, 0.02 pc is about 2 arc seconds on the sky which
is less than the size of the dust tendrils in dense dust regions. This
probably means that the embedded sources only influence the local
structure in regions where the gas and dust are well shielded from the
OB association radiation, such as the tendrils seen behind the capstone
dust areas. On larger scales the radiation pressure from the O star
cluster prevails, and the material enters the general trail of material
swept up by the cluster radiation field.

\section{Column III Sources} \label{sec-c3}

The crescent area at the top of column III contains two stars, MS16S-1
and MS16S-2, both with F205W Vega system magnitudes of 13.2.  The
northern star (Star 1 in Fig.~\ref{fig-c2}) has a slightly redder color
with an F110W Vega magnitude of 17.3 as opposed to 17.0 for the
southern star (Star 2).  Table~\ref{tab-dat} gives the full set of
magnitudes for the two stars.  We assume the red color of these stars
is due to the foreground screen of dust that covers the entire extent
of the three column structure.  Given the similarity of the two stars
it is appropriate to ask if they are physically associated, whether
they formed in column III and whether the crescent shape of the tip of
column III represents a cavity produced by a combination of the
radiation pressure from the two stars and the much stronger radiation
pressure from the OB association.  If the stars are at the 2 kpc
distance of M16 their 2 $\micron$ absolute magnitude is 1.7, assuming
no extinction at 2 \micron.  This would place them at about spectral
type A3 on the main sequence with luminosities on the order of 30
L$_{\odot}$.  If there is extinction at 2 \micron, as there surely must
be, the two stars will have luminosities greater than 30 L$_{\odot}$.

The radius of the crescent in the tip of column 3 is about 10 arc
seconds which is equal to 0.1 pc at the distance of M16. It is doubtful
that stars with a combined luminosity of 60 L$_{\odot}$ would produce a
cavity of 0.1 pc but they would certainly reduce the density in the
immediate region around them.  The radiation from the OB association
would then have an easier time photoevaporating that region, creating
the observed crescent cavity. This may be one mechanism for dividing
the main column into many substructures.  If M16S-1 and 2 are
indeed stars that formed at the tip of the column we are seeing a
snapshot of three stages of star formation in the columns, going from
embedded in column I, to emerging in column II, to completely emerged in
column III.

\section{Conclusions}

There are at least two currently active areas of star formation in the
columns of M16.  Both are near the tops of the columns which face the
cluster of luminous O stars.  Although the optical images show
significant extinction throughout the columns, the infrared images
show that the density of dust and molecular gas is quite low, except at
the location of the embedded sources and at isolated areas along the
columns to the southeast. Far infrared images show another strong
emission area to the southeast, beyond the extent of our images. It
appears that this region of M16 is in its last stage of star formation
and the dissipation of the columns may soon follow as the material in
the few dense clumps is exhausted.  This explains the relatively low
rate of star formation commented on by \citet{pil98}.  An alternative is
that further star formation will be triggered in the columns as the
photocompressed region moves down the length of each column.

The dust and molecular gas in the columns away from the dust structures
has a high enough density to be opaque to optical emission but not
enough to produce high amounts of current star formation.  CO emission
along the columns indicates that the amount of dust is sufficient to
shield some molecular gas from dissociation.  The dense gas and dust at
the locations of the two embedded sources act as capstones, shadowing
much of the column from the ionizing radiation of the O star cluster.
The O star cluster, however, has a significant angular extent, and
ionizing radiation does fall on the surface of the columns.  The
detailed interaction with the surface is described by \citet{hes96}.

The lack of ionizing radiation from M16ES-1 indicates that it is
either a cluster of relatively low mass stars or an object or objects
that have not yet evolved to the Zero Age Main Sequence. The line
emission in atomic hydrogen is due to
photo-ionization and photo-excitation by radiation from the O star
cluster. The morphology of the dust near the embedded sources suggests
that they have a role in determining the local structure.  The
elementary calculations of Sec.~\ref{sec-loc} indicate that the sources
have sufficient strength to affect their surroundings, but only in
regions well shielded from the radiation from the OB association.  The
restriction of the dust regions to a relatively narrow cone that points
in the direction of the O star cluster shows that the cluster radiation
plays the primary role in the dust and gas dynamics at the site of the
embedded sources.

High resolution camera 2 images show two bright stars of very
similar magnitude and color within the concave depression in the tip of
column III.  These may be a chance superposition of local background
stars on this location.  The stars, however, may have formed in the tip
of the column.  If so the stars may have reduced the density in their
surroundings enough to allow the photoevaporation from the O star
cluster to be much more efficient in this area, forming the current
concave depression in the column tip.  The three columns would then
show a sequence of star formation stages from embedded in column I,
emerging in column II, and fully emerged in column III.

\section{Acknowledgements}

This work is supported in part by NASA grant NAG 5-3042.  This letter
is based on observations with the NASA/ESA Hubble Space Telescope,
obtained at the Space Telescope Science Institute, which is operated by
the Association of Universities for Research in Astronomy under NASA
contract NAS5-26555.  We would like to acknowledge the very helpful
comments of the referee Glenn White.

\clearpage

\begin{deluxetable}{cccccccccc}
\tabletypesize{\scriptsize}
\tablecaption{Source Fluxes, Vega Magnitudes, Luminosities and Positions \label{tab-dat}}
\tablewidth{0pt}
\tablehead{
\colhead{Source} & \colhead{F110W} & \colhead{F110W}  & \colhead{F160W}   
& \colhead{F160W} & \colhead{F222M \tablenotemark{a}} & \colhead{F222M \tablenotemark{a}} & \colhead{Lum.} & \colhead{R.A.} & \colhead{Dec.} \\ \colhead{ } & \colhead{mJy} & \colhead{mag} & \colhead{mJy} & \colhead{mag} & \colhead{mJy} & \colhead{mag} & \colhead{L$_{\sun}$} & \colhead{J2000} & \colhead{J2000}
}
\startdata
M16ES-1 & 0.85 & 15.8 & 4.3 & 13.5 & 25. & 11.1 & 200 & $18\fh$ $18\fm$ $50.29\fs$ & $-13\arcdeg$ $48\arcmin$ $55.2\arcsec$ \\
M16ES-2 & 0.3 &  16.9 & 0.4 & 16.1 & 0.66 & 15.0 & 20 & $18\fh$ $18\fm$ $48.64 \fs$ & $-13\arcdeg$ $49\arcmin$ $50.9\arcsec$ \\
Star 1 & 0.21 & 17.3 & 8.1 & 12.8 & 3.8 & 13.2 & 30 & $18\fh$ $18\fm$ $48.75 \fs$  & $-13\arcdeg$ $50\arcmin$ $38.6\arcsec$ \\
Star 2 & 0.27 & 17.0 & 9.0 & 12.7 & 3.9 & 13.2 & 30 & $18\fh$ $18\fm$ $48.79\fs$  & $-13\arcdeg$ $50\arcmin$ $40.4\arcsec$ \\
\enddata
\tablenotetext{a}{This is F205W for the last two entries}

\end{deluxetable}

\clearpage
\begin{figure}


\caption{Infrared image of the observed regions of Column I and II.
Blue, green and red in the image correspond to the F110W, F160W, and
F222M NICMOS filters. The stretch is square root and has been adjusted
to enhance the visibility of the sources.  The blue emission
surrounding the columns is most likely Paschen $\beta$ and He 10830
that lie in the F110W filter. All portions of the image have the same
stretch. North is up and east is to the left as shown by the compass
points. Each square field of view is 51\arcsec.}

\label{fig-ir}
\end{figure}

\clearpage
\begin{figure}


\caption{Infrared image of the observed region in Column III with
camera 2. The image uses the F110W, F160W, and F205W NICMOS filters,
represented by blue, green, and red.  The stretch is linear and has
been adjusted to enhance the visibility of the sources.  As in
Fig.~\ref{fig-ir} the blue emission surrounding the column is most
likely Paschen $\beta$ and He 10830 that lie in the F110W filter.  The
red round region to the north of the dark column II is thermal emission
from the coronagraphic hole in camera 2.  It is not perfectly round
since it is the image from 4 offset points in the dither pattern.  When
the images are added with the stars aligned the coronagraphic hole
images are slightly misaligned.  The square field of view is 17\arcsec.}

\label{fig-c2}
\end{figure}

\clearpage
\begin{figure}


\caption{The same region of the M16 columns in the optical from WFPC2
and the three NICMOS infrared images at 1.1, 1.6 and 2.2 \micron.  All
of the images have a linear stretch.  The regions marked a, b, c and d
in the F160W image indicate the regions used in the extinction and
density analysis of column I described in section~\ref{ssec-ex}. As in
fig.~\ref{fig-ir} each square field of view is 51\arcsec.}

\label{fig-3colors}
\end{figure}

\clearpage
\begin{figure}


\caption{Four images of the same region at the top of column 1 in
Paschen $\alpha$, H $\alpha$, H$_2$, and $2.2 \micron$ continuum.  All
regions have the same scale and orientation and the intensity scale is
linear in ADUs per second.  Subtraction of the continuum image can
produce negative values in the line emission images since the psfs of
the line and continuum are slightly different due to their differing
wavelengths. This produces dark holes at the positions of bright
stars. The square field of view is 51\arcsec.}

\label{fig-lines} \end{figure}


\clearpage



\begin{thebibliography}{}

\bibitem[Allen et al. (1999)]{all99} Allen, L. E. et al. 1999, \mnras, 304, 98

\bibitem[Black and van Dishoeck (1987)]{bl87} Black, J. H. \& van Dishoeck, 
	E. F. 1987, \apj, 322, 412.

\bibitem[Hester et al. (1996)]{hes96} Hester, J. J., et al. 1996, \aj, 111, 2349

\bibitem[Hester (1997)]{hes97} Hester, J. J. 1997 in Star Formation Near and
	Far, ed. S. S. Holt \& L. G. Mundy, (New York: AIP Press), 143

\bibitem[Hester (2002)]{hes02} Hester, J. J. 2002, in preparation.

\bibitem[Hillenbrand et al. (1993)]{hil93} Hillenbrand, L. A., Massey, P.,
	Strom, S. E., \& Merrill, K. M. 1993, \aj, 106, 1906

\bibitem[Lada, Alves and Lada (1999)]{lad99} Lada, C. J., Alves, J., \& 
	Lada, E. A. 1999, \apj, 512, 250

\bibitem[Lombardi and Alves (2001)]{lom01} Lombardi, M. \& Alves, J. 2001, 
	\aap, in press

\bibitem[Lytle et al. (1999)]{sto99} Lytle, D., Stobie, E., Ferro, A.,
	\& Barg, M. 1999, in ASP Conf. Ser. 172, Astronomical Data Analysis
	Software and Systems VIII, ed. D. Mehringer, R. Plante, \&
	D. Roberts, (San Francisco: ASP), 445

\bibitem[McCaughrean (1997)]{mcc97} McCaughrean, M. 1997, in IAU Symposium 182,
	Herbig-Haro Flows and the Birth of Low Mass Stars, ed. B. Reipurth and
	C. Bertout (Dordrecht: Kluwer), 551

\bibitem[Pilbratt et al. (1998)]{pil98} Pilbratt, G. L., Altieri, B., Blommaert
	A. D. L., Fridlund, C. V. M., Tauber, J. A., \& Kessler, M. F. 1998,
	\aap, 333, L9

\bibitem[Rieke and Lebofsky (1985)]{rie85} Rieke, G. H. \& Lebofsky, M. J. 1985,
	\apj, 288, 618

\bibitem[Seward (1999)] {sew99} Seward, F. D. 1999, in 
	Allen's Astrophysical Quantities, 4th Edition, ed  A. N. Cox , 
	(New York: Springer-Verlag), 197
	
\bibitem[Smith et al. (2002)]{smth02} Smith, B. A., et al. in preparation

\bibitem[Stetson (2001)]{stet01} Stetson, P. B. 2001, Users Manual for 
	DAOPHOT II

\bibitem[Thompson (1984)]{th84} Thompson, R. I. 1984, \apj, 283, 165

\bibitem[Thompson et al. (1999)]{thm99} Thompson, R. I., Storrie-Lombardi,
	L. J., Weymann, R. J., Rieke, M. J., Schneider, G., Stobie, E.,
	Lytle, D. 1999, \aj, 117, 17

\bibitem[White et al. (1999)]{wh99} White et al. 1999, \aap, 342, 233

\end{thebibliography}
\end{document}